\documentclass[aps,prb,reprint,notitlepage,showkeys,superscriptaddress,floatfix,a4paper]{revtex4-1}
\usepackage{graphicx}
\usepackage{epsfig}
\usepackage[update,prepend]{epstopdf}
\usepackage{amssymb}
\usepackage{amsmath}
\usepackage{graphics}
\usepackage{latexsym}
\usepackage{color}
\newcommand{\la}{\left\langle}
\newcommand{\ra}{\right\rangle}
\def \figwidth {\columnwidth}

\begin{document}

\title{Entropic Unmixing in Nematic Blends of Semiflexible Polymers}
\author{Andrey Milchev}
\affiliation{Institute for Physical Chemistry, Bulgarian Academia of Sciences, 1113 Sofia, Bulgaria}
\affiliation{Institute of Physics, Johannes Gutenberg University Mainz, Staudingerweg 7, 55128 Mainz,
Germany}

\author{Sergei A. Egorov}
\affiliation{Department of Chemistry, University of Virginia, Charlottesville, VA 22901, USA}
\affiliation{Institute of Physics, Johannes Gutenberg University Mainz, Staudingerweg 7, 55128 Mainz,
Germany}

\author{Jiarul Midya}
\affiliation{Institute of Physics, Johannes Gutenberg University Mainz, Staudingerweg 7, 55128 Mainz,
Germany}

\author{Kurt Binder}
\affiliation{Institute of Physics, Johannes Gutenberg University Mainz, Staudingerweg 7, 55128 Mainz,
Germany}

\author{Arash Nikoubashman}
\affiliation{Institute of Physics, Johannes Gutenberg University Mainz, Staudingerweg 7, 55128 Mainz,
Germany}
\email{anikouba@uni-mainz.de}

\date{\today}


\begin{abstract}
Binary mixtures of semiflexible polymers with the same chain length but different persistence lengths separate into two coexisting different nematic phases when the osmotic pressure of the lyotropic solution is varied. Molecular Dynamics simulations and Density Functional Theory predict phase diagrams either with a triple point, where the isotropic phase coexists with two nematic phases, or a critical point of unmixing within the nematic mixture. The difference in locally preferred bond angles between the constituents drives this unmixing without any attractive interactions between monomers.
\end{abstract}

\maketitle
Favorable materials properties can be achieved by processing blends from chemically different constituents, {\it e.g.}, addition of poly(vinyl chloride) for permanent plasticization\cite{robeson:poly:1984} or mixing of poly(phenylene ether) resins and poly(styrene) for materials with high heat resistance and low density.\cite{hay:poly:1998} Depending on the application, homogeneous or heterogeneous polymer blends are desired, and their phase behavior has been mapped for a wide range of different polymer chemistries.\cite{robeson:poly:1984, dutta:poly:1990, hay:poly:1998, paul:book:2000, konigsveld:book:2001} Extensive theoretical work has also been performed to rationalize and predict the phase behavior of polymer blends, mainly focusing on the enthalpic interactions between the different monomeric units.\cite{paul:book:2000, konigsveld:book:2001, rubinstein:book:2003} Semiflexible macromolecules, which are almost rigid over the scale of the persistence length along the chain backbone but flexible on larger scales,\cite{rubinstein:book:2003, donald:book:2006} are of particular interest due to their ubiquity in biological systems\cite{broedersz:rev:2014, koester:bio:2015} and their anisotropic physical properties in the liquid crystalline state. In lyotropic solutions or blends of semiflexible polymers, entropic effects alone can drive a transition from an isotropic (i) to a nematic (n) phase, which is accompanied by a distinct change of the materials elastic properties. This (macroscopic) phase behavior strongly depends on the (microscopic) bending stiffness of the macromolecules, and understanding these properties is a challenge for statistical mechanics due to the numerous disparate length-scales involved.\cite{degennes:book:1979, rubinstein:book:2003, donald:book:2006, kato:poly:2018, allen:mp:2019, binder:jpm:2020} The ordering of semiflexible polymers is also central for various applications, and therefore controlling the polymer stiffness has become a very active area of research.\cite{chuang:prl:2017, baun:mm:2020}

Experimentally, it is challenging to unambiguously differentiate between the (enthalpic) contributions due to polymer chemistry and the (entropic) contributions due to polymer stiffness. The effect of polymer stiffness on unmixing has been neglected in most theoretical descriptions, presumably because it does not play a role in the standard Flory-Huggins (FH) mean field theory.\cite{rubinstein:book:2003, flory:book:1953} A mathematically elegant theory for thermotropic solutions and blends of semiflexible polymers has been developed by Liu and Fredrickson,\cite{liu:mm:1993} using a Landau expansion in terms of two order parameters. For the solution, one order parameter is the deviation of the local volume fraction of the polymer from its average, while the other is the local nematic tensor order parameter. As Landau theory is based on a power series expansion of the order parameters, it is applicable when the order parameters are sufficiently small, but becomes unsuitable deep in the nematic phase where the order parameters approach their saturation values.\cite{degennes:clc:1984} Hence, Liu and Fredrickson focused only on the phase behavior in the vicinity of i-i and i-n phase transitions. In their treatment, these transitions are driven by enthalpic interactions throughout, postulating a standard isotropic FH parameter to be present, as well as a Maier-Saupe-like term\cite{degennes:book:1993} driving the nematic ordering in the thermotropic solution.\cite{liu:mm:1993} The analogous transitions for the (incompressible) blend were also briefly discussed, but the situation deep in the nematic phase could not be addressed. Therefore, the majority of previous work focused on mixtures of fully flexible polymers with hard rods,\cite{liu:mm:1996, yang:jpp:2001, oyarzun:jcp:2015, ramirez:poly:2017} or on mixtures of polymers with little stiffness in the isotropic phase.\cite{kozuch:poly:2016, fredrickson:mm:1994} Holyst and Schick suggested the existence of an n-n coexistence region for mixtures of two types of strictly rigid rods with comparable length, driven by enthalpic interactions.\cite{holyst:jcp:1991} The existence of thermotropic n-n unmixing was also found by experiments on mixtures of side-chain liquid-crystalline polymers and small molecule liquid crystals.\cite{chiu:mm:1996} Note, however, that again the unmixing was driven by the standard enthalpic FH parameter, and {\it not} by the differences in chain stiffness alone.

The present work fills this gap by elucidating the properties of strictly lyotropic solutions of mixtures of two semiflexible polymers (A and B) with strong stiffness, such that each pure component exhibits an i-n transition with increasing osmotic pressure $P$. While unmixing of polymers in solution is also rather common when they exhibit a large disparity in chain length, $N_{\rm A} \gg N_{\rm B}$,\cite{dennison:jcp:2011} we focus here on the limiting case of identical chain length, $N_{\rm A} = N_{\rm B} = N$, where only their persistence lengths differ $\ell^{\rm A}_{\rm p} \neq \ell^{\rm B}_{\rm p}$. We employ Molecular Dynamics (MD) simulations and Density Field Theory (DFT) calculations of a coarse-grained bead-spring model, in which the polymer stiffness is controlled through a bending potential with interaction strength $\kappa$ (see Model section and ESI for details). Choosing a binary mixture of polymers, differing only by their stiffness parameters $\kappa_{\rm A}$ and $\kappa_{\rm B}$, respectively, we explore the phase behavior as functions of $P$ and the mole fraction of B chains, $X_{\rm B}$ ($X_{\rm A} = 1-X_{\rm B}$). We study polymers with $16 \leq \kappa_{\rm A} \leq 128$ at fixed $N=32$ and $\kappa_{\rm B} = 128$, which exhibit i-n transitions with increasing $P$ in lyotropic solution.\cite{egorov:sm:2016, egorov:prl:2016, milchev:jcp:2018, milchev:cms:2019} We show that for low $P$ a wide two-phase coexistence region between isotropic and nematic phases exists, with rather different monomer densities $\rho$ in the coexisting phases. When the ratio $\kappa_{\rm B}/\kappa_{\rm A}$ exceeds a critical value, a triple point occurs  at $P = P_{\rm t}$. For $P > P_{\rm t}$, two nematic phases n$_1$ and n$_2$ coexist, one rich in A chains and the other rich in B chains. For smaller $\kappa_{\rm B}/\kappa_{\rm A}$ above the i-n coexistence region, a homogeneously mixed nematic phase occurs, which splits into an n$_1$-n$_2$ coexistence region only at higher pressures. We elucidate the molecular origin which drives this phase separation between very similar nematic phases. 

\begin{figure}[htbp]
	\includegraphics[width=\figwidth]{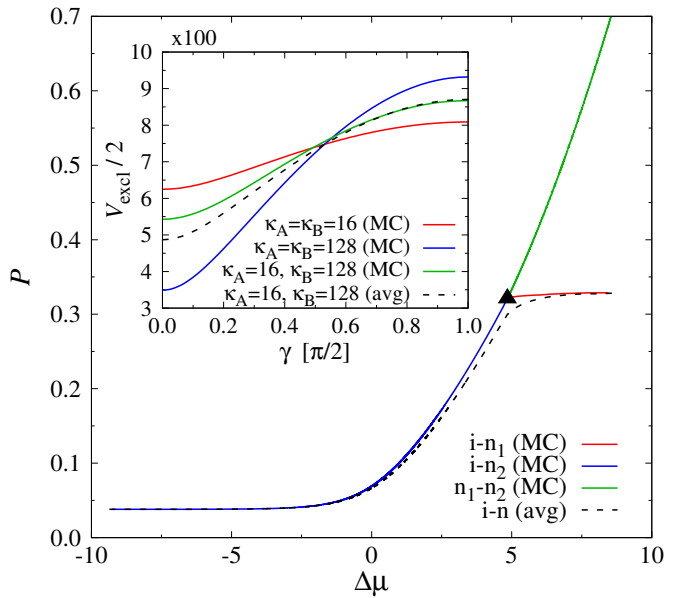}
	\caption{Phase diagram from DFT in the plane of intensive thermodynamic variables pressure $P$ {\it vs.} chemical potential difference $\Delta \mu$ between the A and B species, for $\kappa_{\rm A}=16$ and $\kappa_{\rm B}=128$. Results shown for calculations where $V_{\rm excl}^{\alpha \beta}(\gamma)$ was computed from MC simulations with $\kappa_{\rm A}=16$ and $\kappa_{\rm B}=128$, and where the average $V^{\rm avg}_{\rm excl}(\gamma) = \left[V^{\rm AA}_{\rm excl}(\gamma) + V_{\rm excl}^{\rm BB}(\gamma)\right]/2$ was used. Inset shows corresponding $V_{\rm excl}$ terms as functions of angle $\gamma$.}
	\label{fig:vexc}
\end{figure}

Can one predict the behavior of the mixed system from knowledge on the pure components alone? To answer this question, let us first consider the excluded volume interactions between two semiflexible chains at angle $\gamma$ between their molecular axes, $V_{\rm excl}^{\rm AB}(\gamma)$ (see ESI for technical details). In the inset of Fig.~\ref{fig:vexc} we compare the actual $V_{\rm excl}^{\rm AB}(\gamma)$ with the average $V_{\rm excl}^{\rm avg}(\gamma) = [V_{\rm excl}^{\rm AA} + V_{\rm excl}^{\rm BB}]/2$, finding that the data are indistinguishable for large $\gamma$, but differ for small $\gamma$. These small differences are, however, crucial as the resulting phase diagrams in Fig.~\ref{fig:vexc} demonstrate: the approximation $V_{\rm excl}^{\rm avg}$ predicts correctly the i-n phase boundary over almost the full range of chemical potential difference $\Delta \mu$, but fails to capture the existence of a triple point and n$_1$-n$_2$ transition line ending there. In fact, the difference $V^{\rm AB}_{\rm excl} - V_{\rm excl}^{\rm avg}$ plays the role of an FH $\chi$-parameter causing the n$_1$-n$_2$ phase separation.

The calculated $V_{\rm excl}^{\rm AB}(\gamma)$ suffer from statistical errors, and our DFT calculations do not include correlations between monomer positions due to dense packing of chains explicitly (see ESI for details). Thus it is crucial to test DFT by MD work. Both in MD and in experiment, $\Delta \mu$ is not accessible, and hence phase diagrams using $X_{\rm B}$ rather than $\Delta \mu$ are studied. For determining the phase diagrams in MD, we simulated $\mathcal{N}/N = 6,144$ chains in an elongated box with $L_y = 32$ and $L_z = 64$, where we varied the length $L_x$ to achieve the desired monomer number density, $\rho$. Starting configurations were prepared as ordered arrays of rods (parallel to the $z$-direction), which were initially separated into pure A and B phases so that all A chains (B chains) were located at $x<0$ ($x>0$). Then, phase diagrams have been determined by computing the coexisting densities in the A-rich and B-rich phases after the systems reached equilibrium.

\begin{figure*}[htbp]
	\centering
	\includegraphics[height=10.2cm]{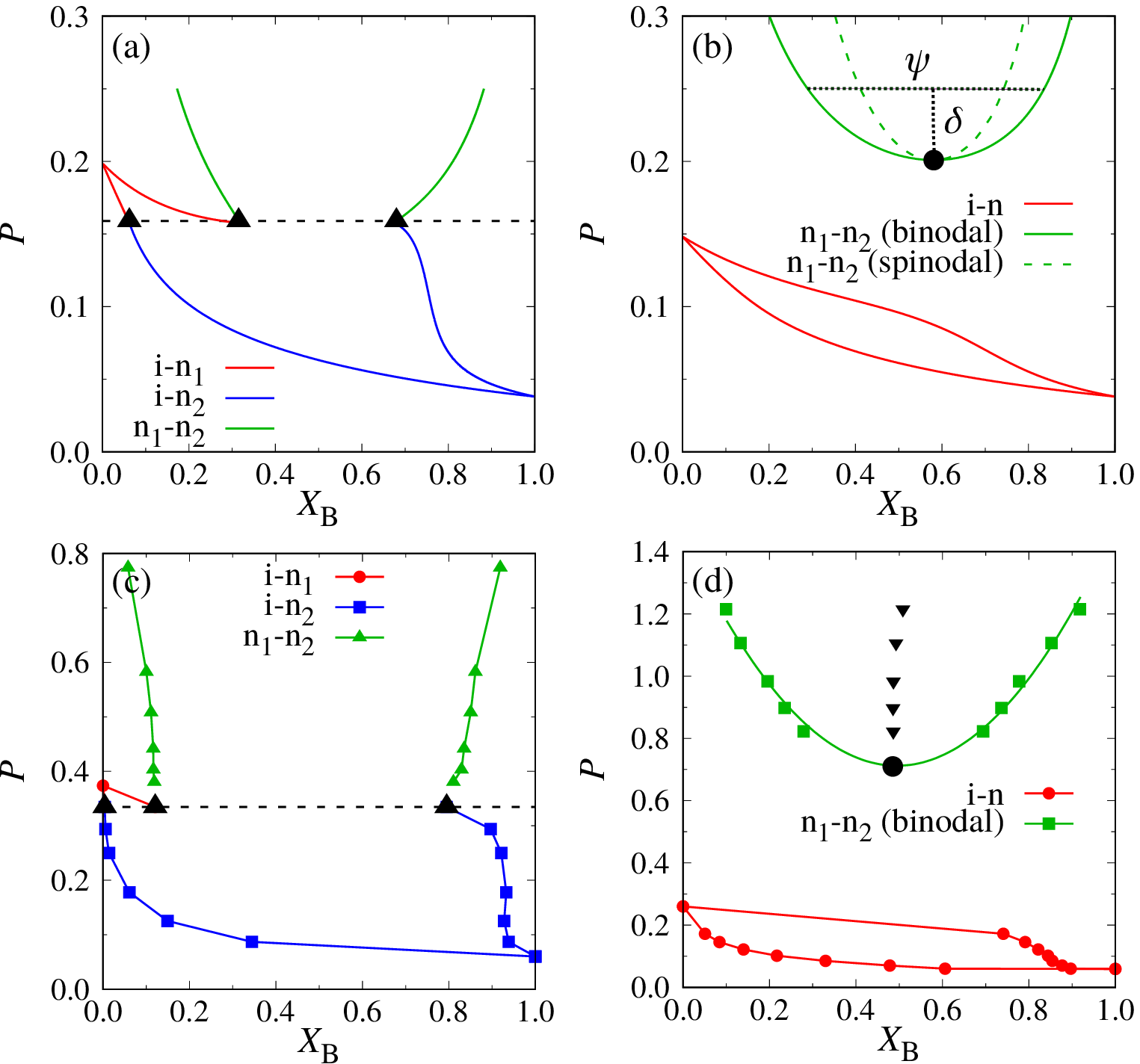}
	\quad
	\includegraphics[height=10.2cm]{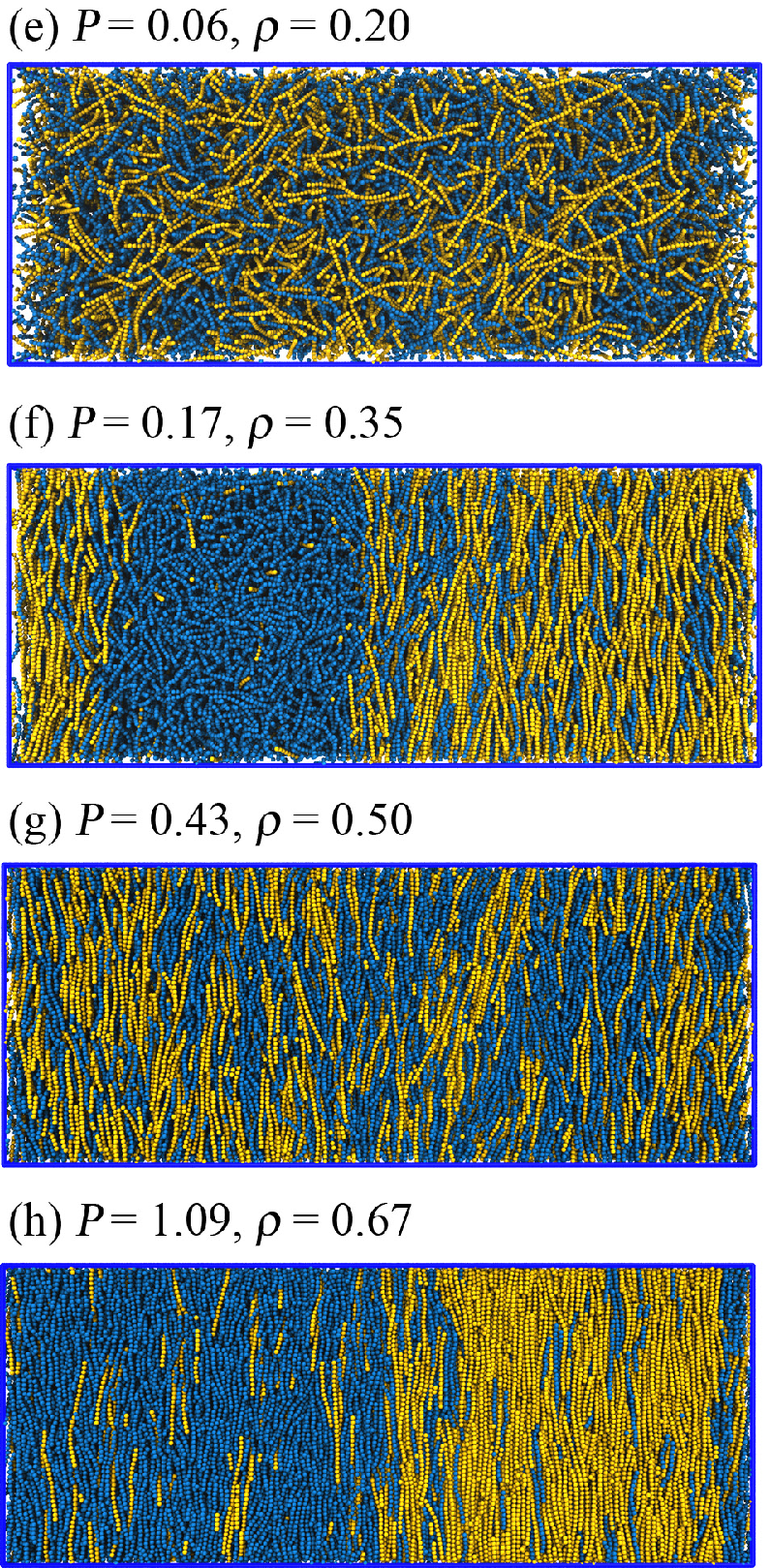}
	\caption{Phase diagrams of binary mixtures of semiflexible polymers in the $P$-$X_{\rm B}$ plane, according to DFT (a) $\kappa_{\rm A} = 20$, (b) $\kappa_{\rm A} = 24$, and MD (c) $\kappa_{\rm A} = 16$, (d) $\kappa_{\rm A} = 20$. In (a,c), the dashed horizontal line corresponds to $P_{\rm t}$, and black triangles indicate triple points. Black dots in (b,d) indicate critical points, and black triangles in (d) show the rectilinear diameter. The relative pressure difference $\delta = (P - P_{\rm c})/P_{\rm c}$ and the concentration difference $\psi = X_{\rm B}({\rm n}_2) - X_{\rm B}({\rm n}_1)$ are indicated in panel (b). (e-h) Simulation snapshots at $X_{\rm B} = 1/2$ for the data shown in (d), with A and B chains colored in blue and yellow, respectively.}
	\label{fig:pd}
\end{figure*}

Figure \ref{fig:pd}(a-d) shows the resulting phase diagrams from DFT and MD in the plane of variables $X_{\rm B}$ and $P$, while Fig.~\ref{fig:pd}(e-h) displays selected snapshots from MD simulations at $\kappa_{\rm A} = 20$. With increasing $P$, first an i-n miscibility gap opens. For a large enough ratio $\kappa_{\rm B}/\kappa_{\rm A}$ it ends at a triple point $P_{\rm t}$, where n$_1$-n$_2$ phase separation takes over. For somewhat smaller $\kappa_{\rm B}/\kappa_{\rm A}$, a region of homogeneous nematic mixture occurs, before n$_1$-n$_2$ unmixing starts at a critical point. For too small $\kappa_{\rm B}/\kappa_{\rm A}$, however, this transition would be preempted by smectic or crystal phases.\cite{milchev:cms:2019} Both methods predict qualitatively similar phase diagrams, but the prediction for the multicritical point $\kappa_{\rm A}^{\rm m}$, where the triple point disappears in favor of an n$_1$-n$_2$ critical point, differ: $\kappa_{\rm A}^{\rm m} = 20.5$ in DFT while MD implies $\kappa_{\rm A}^{\rm m} = 18 \pm 1$.

In our MD simulations, there is no {\it ad hoc} assumption about specific AB-repulsions, all pairs of monomers interact with the same purely repulsive interaction. We follow ideas of Kozuch {\it et al.}\cite{kozuch:poly:2016} to show how the mismatch of stiffness can yield an effective FH $\chi$-parameter nevertheless: The free energy of the mixture can be expressed as
\begin{equation}
	\frac{F}{Nk_{\rm B}T} = \frac{1}{N}(X_{\rm A} \ln X_{\rm A} + X_{\rm B} \ln X_{\rm B}) + \chi X_{\rm A} X_{\rm B},
\end{equation}
where the first and second term account for the entropy and enthalpy change as a result of mixing, respectively, $k_{\rm B}$ is Boltzmann's constant, and $T$ is the temperature of the system. The non-ideal mixing term $\Delta F_{\rm exc}/(Nk_{\rm B}T) = \chi X_{\rm A}X_{\rm B}$ can also (for $X_{\rm B} = 1/2$) be written as $\Delta F_{\rm exc}(\kappa_{\rm A}, \kappa_{\rm B})=F_{\rm AB}(\kappa_{\rm A}, \kappa_{\rm B})-[F_{\rm A}(\kappa_{\rm A}) +F_{\rm B} (\kappa_{\rm B})]/2$. Then we can estimate $\Delta F_{\rm exc}$ through thermodynamic integration of the difference in bending energies\cite{kozuch:poly:2016}
\begin{equation}
	\Delta F_{\rm exc}/N = \frac{1}{2} \int^{\kappa_{\rm B}}_{\kappa_{\rm A}} {\rm d}\kappa \Delta \cos(\theta_{ijk}) ,
	\label{eq3}
\end{equation}
where $\Delta \cos(\theta_{ijk}) = \la \cos(\theta_{ijk})\ra_{\rm AB}^{\rm A} - \la \cos(\theta_{ijk})\ra_{\rm A}^{\rm A}$ is the difference in bending angles of an A chain in an AB-environment and in a pure A phase. We sampled $\Delta \cos(\theta_{ijk})$ through additional MD simulations at fixed monomer density $\rho=0.42$ and composition $X_{\rm B}=1/2$. We systematically reduced $\kappa_{\rm A}$ from $\kappa_{\rm A}=\kappa_{\rm B}=128$ to $\kappa_{\rm A} = 18$ to stay in the mixed regime of the phase diagram. Simulations were performed in a cubic simulation box ($L_x=L_y=L_z=64$), with the polymers initialized as fully mixed arrays of ordered rods.

\begin{figure}[htbp]
	\includegraphics[width=\figwidth]{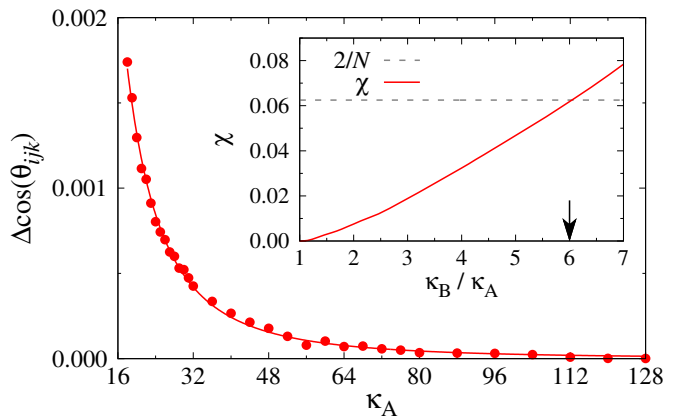}
	\caption{Difference in bending angles $\Delta \cos(\theta_{ijk})$ {\it vs.} $\kappa_{\rm A}$ at constant monomer density $\rho = 0.42$. The solid line is intended as a guide to the eye only. The inset shows the resulting $\chi$-parameter plotted vs. $\kappa_{\rm B}/\kappa_{\rm A}$. The horizontal line shows the mean field prediction for criticality, $\chi_{\rm c} = 2/N$, and the black arrow indicates the position where the two curves cross.}
	\label{fig:FH}
\end{figure}

Figure~\ref{fig:FH} shows the resulting $\Delta \cos(\theta_{ijk})$ as a function of $\kappa_{\rm A}$. It is seen that when $\kappa_{\rm A}$ is not much smaller than $\kappa_{\rm B}$, the difference in bending angles is small but rises steeply for $\kappa_{\rm B}/\kappa_{\rm A} > 4$, and criticality\cite{degennes:book:1979, rubinstein:book:2003, flory:book:1953} ($\chi_{\rm c} = 2/N$) is reached here for $\kappa_{\rm B}/\kappa_{\rm A} \approx 6$, i.e. $\kappa_{\rm A}^{\rm m} \approx 21.3$. The pressure for $\rho = 0.42$ ($P \approx 0.29$) clearly falls into the i-n two-phase coexistence region of $\kappa_{\rm A} = 16$ (Fig.~\ref{fig:pd}c). For $\kappa_{\rm A} = 20$, however, the critical pressure $P_{\rm c} \approx 0.7$ (Fig.~\ref{fig:pd}d) corresponds to $\rho_{\rm c} \approx 0.58$, and hence the system with $\rho = 0.42$ still is in the region of the fully mixed nematic phase. We expect that the FH $\chi$-parameter near criticality is proportional to $P$ (or $\rho$, respectively), which might explain the small difference for the estimated $\kappa_{\rm A}^{\rm m}$.

A consideration based on the Ginzburg criterion\cite{ginzburg:rus:1961, alsnielsen:ajp:1977} suggests that n$_1$-n$_2$ unmixing has an extended meanfield-like critical region for large enough $N$, before the universal Ising critical behavior\cite{yeomans:book:1992, povodyrev:pa:1999} sets in close enough to $P_{\rm c}$. For the Ginzburg criterion, one needs to estimate how the correlation lengths $\xi$ scale within mean field theory. Here, the correlations are very anisotropic due to the nematic order in the system. Recall that the random phase approximation \cite{degennes:book:1979, rubinstein:book:2003} relates the correlation lengths $\xi$ to the radii of gyration as
\begin{equation}\label{eq4}
	\xi_{\parallel, \perp}^2 = \left[\left(X_{\rm B}X_{\rm A}\right)^{-1}-2N \chi\right]^{-1} \left[\frac{\la R_g^2 \ra_{\parallel, \perp}^{\rm B}}{X_{\rm B}}+ \frac {\la R_g^2\ra_{\parallel, \perp}^ {\rm A}}{X_{\rm A}} \right] ,
\end{equation}
where the subscripts $\parallel$ and $\perp$ indicate the linear dimensions along the nematic director and perpendicular to it. Each chain occupies approximately a cylindrical volume $N\ell_{\rm b}\pi R^2$ with $R^2=1/(\pi\ell_{\rm b}\rho)$, and the critical density is about three times the density of the onset of nematic order, $\rho_{\rm n}$ (Fig.~\ref{fig:pd}). Since $\rho_{\rm n}$ scales like $1/N$ for very stiff polymers, we predict $R \propto \sqrt{N}$ for large $N$. The gyration radius in the nematic phase is of the same order as $R$, and hence the correlation length prefactors scale as $\xi_\perp \propto \sqrt{N}$ and $\xi_\parallel \propto N$. To the best of our knowledge, such a strongly anisotropic type of critical behavior is not yet known for other systems.

While for blends of flexible polymers the relation $\chi_{\rm c} \propto 1/N$ could be verified by varying $N$ over a wide range \cite{gehlsen:prl:1992, deutsch:epl:1992}, and the Ising to mean field crossover studied,\cite{deutsch:epl:1993} a corresponding study for n$_1$-n$_2$ unmixing is very challenging. While in the isotropic case of flexible blends the correlation volume scales with $N$ as $N^{3/2}$, for n$_1$-n$_2$ unmixing the correlation volume scales as $\xi_\parallel \xi_\perp^2 \propto N^2 \delta^{-3/2}$, where $\delta = (P - P_{\rm c})/P_{\rm c}$. The concentration difference $\psi = X_{\rm B}({\rm n}_2) - X_{\rm B}({\rm n}_1)$ scales as $\delta^{1/2}$, and the corresponding mean-square fluctuation per unit volume as $N/\delta$.\cite{binder:adv:1994} Hence, the average mean-square fluctuation per correlation volume scales $\delta^{1/2}/N$, which is much smaller than $\psi^2$ if $\delta \gg N^{-2}$. Therefore, the mean field critical exponents are self consistent, except very close to $P_{\rm c}$ where Ising criticality takes over. This Ginzburg criterion only concerns exponents, it does not imply that the relation $\chi_{\rm c} = 2/N$ is accurate.\cite{binder:adv:1994}

In these binary mixtures, the chain properties differ from their pure counterparts: e.g. in the n$_1$-n$_2$ coexistence region for $\kappa_{\rm A} = 16$ at $P=0.44$ ($\rho=0.50$), the nematic bond order parameters are $S_{\rm A} \approx 0.66$, $S_{\rm B} \approx 0.85$ in the A-rich phase, and $S_{\rm A} \approx 0.80$, $S_{\rm B} \approx 0.93$ in the B-rich phase, whereas the corresponding values of the pure phases are $S_{\rm A} \approx 0.61$, $S_{\rm B} \approx 0.93$. Also the components of the radius of gyration tensor show that the chains must accommodate to their environment: In the direction parallel to the nematic director, we found $\la R_g^2\ra^{\rm A}_\parallel \approx 61$ in the A-rich phase and $\approx 69$ in the B-rich phase, whereas the component perpendicular to the director was $\la R_g^2\ra^{\rm A}_\perp \approx 6.6$ in the A-rich phase but only $\approx 2.6$ in the B-rich phase. The less stiff A-chains need more space in the transverse direction, and this misfit drives the n$_1$-n$_2$ phase separations, and causes also an appreciable density difference between the coexisting phases ($\rho_{\rm A-rich} \approx 0.471$ and $\rho_{\rm B-rich} \approx 0.529$ in the above example).

Based on DFT and MD model calculations, we predict that blends of rather stiff semiflexible polymers show nematic-nematic unmixing, if the stiffness disparity is large enough. While standard theories \cite{flory:book:1953, paul:book:2000, konigsveld:book:2001, rubinstein:book:2003} attribute polymer unmixing solely to differences in pairwise monomer interactions, we show that unmixing can also be driven by stiffness mismatch, even if the pairwise monomer interactions are still all strictly identical. This transition is driven by the geometric mismatch of the rod-like chains (less stiff polymers need more space in the directions transverse to the director). Critical phenomena have a very unusual anisotropic character and cross over to mean field type behavior for very stiff and very long chains. In future research, a simulation study of early stages of spinodal decomposition of blends quenched from the isotropic phase into a two phase region would be illuminating, but requires an even larger numerical effort due to the need of averaging over multiple runs.

\section{Methods}
A detailed discussion of the employed methods is included in the ESI, and we will provide here only the most important details. The MD simulations use a bead-spring model where each monomer has diameter $\sigma$ and mass $m$. Excluded volume interactions between monomers from the A and B chains are identical, and are taken into account by the Weeks-Chandler-Andersen (WCA) potential.\cite{weeks:jcp:1971} Successive beads along the chains are bound together by the finitely extensible nonlinear elastic (FENE) potential with bond length $\ell_{\rm b} \approx 0.97\,\sigma$.\cite{grest:pra:1986} Bending stiffness is included through
\begin{equation}
	U_{\rm bend}(\theta_{ijk})= \kappa \left[1-\cos\left(\theta_{ijk}\right)\right] \approx \frac{\kappa}{2} \theta ^2_{ijk} ,
	\label{eq1}
\end{equation}
where $\kappa$ controls the interaction strength, and $\theta_{ijk}$ is the angle between the bonds connecting consecutive monomers $i$ to $j$ and $j$ to $k$. The persistence length of the polymers is then $\ell_{\rm p} \approx \ell_{\rm b} \kappa/(k_{\rm B}T)$ for $\kappa \gtrsim 2\,k_{\rm B}T$ and at densities below the isotropic-nematic transition,\cite{milchev:jcp:2018} as expected from the equipartition theorem. Thus, using two different constants $\kappa_{\rm A}$, $\kappa_{\rm B}$ in Eq.~(\ref{eq1}) is the only distinction between the two polymer species.

MD runs were carried out in the $\mathcal{N}VT$ ensemble, with $\mathcal{N}$ being the total number of monomers in the system. The interaction strength of the WCA potential, $\varepsilon$, the bead diameter, $\sigma$, and the monomer mass, $m$, define the units of energy, length, and mass, respectively, in our MD simulations. The intrinsic MD time unit is then $\tau_{\rm MD} = \sqrt{m \sigma^2/\epsilon}$. In the remainder of this manuscript, we omit these units for brevity. The temperature was held constant at $T=1.0$ by a Langevin thermostat with friction constant $\zeta = 1.0$, and a time step $\Delta t= 0.005$ was used for integrating the equations of motion. All runs were carried out with the HOOMD-blue software \cite{anderson:jcp:2008, glaser:cpc:2015} on graphics processing units. The systems were equilibrated for $2 \times 10^8$ to $10^9$ time steps, depending on monomer density, and measurements were then taken over a similar additional period of time.

The DFT calculations use the same bending potential Eq.~(\ref{eq1}), but the chains are represented by a slightly different model using tangent hard spheres of diameter $\sigma$, so that $\ell_{\rm b} = \sigma$. The free energy functional contains an excess term $F_{\rm exc}$ depending on the excluded volume $V_{\rm excl}(\gamma)$ between two semiflexible polymers at angle $\gamma$ between their molecular axes. This interaction $V_{\rm excl}(\gamma)$ cannot be calculated from first principles, but is estimated from dedicated Monte Carlo (MC) simulations,\cite{fynewever:jcp:1998} see Fig.~\ref{fig:vexc}. With $\mathcal{N}/N$ chains in the considered volume, the excess per chain is 
\begin{equation}
	\frac{F_{\rm exc}}{(\mathcal{N}/N) k_{\rm B}T}= \frac{1}{2} \frac{\rho}{N} \int {\rm d} \omega \int {\rm d} \omega' f(\omega)f(\omega') V_{\rm excl}(\gamma(\omega, \omega'))
	\label{eq2}
\end{equation}
where $f(\omega)$ is the orientational distribution function of a bond, and $\omega \equiv (\vartheta, \varphi)$ are the polar angles. The prefactor $\rho/N$ needs to be enhanced by appropriate rescaling \cite{egorov:sm:2016}, and for $V_{\rm excl}(\gamma)$ we now need to distinguish between A-A, B-B and A-B pairs (Fig.~\ref{fig:vexc}). All $V_{\rm excl}^{\alpha \beta} (\gamma)$ are strongly repulsive interactions and of order $N^2$.

\section*{Acknowledgments}
We thank the German Research Foundation (DFG) for support under project numbers BI 314/24-2, NI 1487/4-2, NI 1487/2-1, and NI 1487/2-2. S.A.E. thanks the Alexander von Humboldt Foundation for support. The authors gratefully acknowledge the computing time granted on the supercomputer Mogon (hpc.uni-mainz.de). A.M. thanks the COST action No. CA17139, supported by COST (European Cooperation in Science and Technology) and its Bulgarian partner FNI/MON under KOST-11.

\end{document}